# Search for shortest paths based on a projective description of unweighted graphs

V.A Melent'ev

A.V. Rzhanov Institute of Semiconductor Physics, Siberian Branch of the Russian Academy of Sciences, 13 Lavrentyev aven., Novosibirsk, 630090, Russia.

Contributing authors: melva@isp.nsc.ru;

**Abstract**

The search is based on the preliminary transformation of matrices or adjacency lists traditionally used in the study of graphs into projections cleared of redundant information (refined) followed by the selection of the desired shortest paths. Each projection contains complete information about all the shortest paths from its base (angle vertex) and is based on an enumeration of reachability relations, more complex than the traditionally used binary adjacency relations. The class of graphs considered was expanded to mixed graphs containing both undirected and oriented edges (arcs). A method for representing graph projections in computer memory and finding shortest paths using them is proposed. The reduction in the algorithmic complexity achieved, at the same time, will allow the proposed method to be used in information network applications, scientific and technical, transport and logistics, and economic fields.

# 1 Introduction

The problem of finding shortest paths is central to a wide variety of fields, from computer science to social and transportation networks. Due to the ever-increasing complexity of modern networks, the efficiency of finding shortest paths plays a decisive role [1].

A historical excursus in this field demonstrates the evolution of algorithms from classical algorithms, such as Dijkstra [2], Bellman-Ford [3], A* [4] and Floyd-Warshall [5, 6], to modifications using data structures optimized for them [7]. However, when scaling applications that require real-time processing of graphs with millions of vertices and edges, serious limitations, that make it difficult to apply classical algorithms, arise [8]. Because of these limitations, researchers look for efficiently scalable approaches based not only on optimizing data structures in algorithms, but also on using more efficient graph descriptions.

One such approach is to replace the traditional descriptions of the graph based on binary relations of vertex adjacency with a projective description of it - a system of projections using larger n-ary relations of reachability of the remaining vertices of the graph from the vertices that determine the angles of projections of such a system [9] - [12]. Since search algorithms are essentially combinatorial, the use of a smaller number of larger structural elements should significantly reduce the variability and, consequently, the complexity of algorithms based on them [13].

The main problems of finding shortest paths in graphs are [14]: SSSP (Single-Source Shortest Paths), SDSP (Single-Destination Shortest Paths), SPSP (Single-Pair Shortest Path) and APSP (All-Pairs Shortest Paths). In this paper, we will touch upon each of these problems related to an unweighted mixed graph without loops and multiple edges, using the example of the SSSP problem. However, the information provided, in this case, will be sufficient to find ways in other listed formulations based not on traditional matrix or list adjacency descriptions of the graph, but on a higher-level description of the graph by its projections.

## 2 Projection of an unweighted mixed graph

This section provides the necessary explanations and definitions for the original terminology used in the work. Recall that we are talking here about simple (without loops and multiple edges) mixed graphs. A mixed graph consists of a finite set of $V$-vertices and sets of disordered and ordered pairs of different vertices. Any ordered pair $(u, v)$ is called arc, or oriented edge. The arc $(u, v)$ is directed from vertex $u$ to vertex $v$ and is incident to both of these vertices, while vertex $u$ is adjacent to vertex $v$, and $v$ is adjacent from $u$ [15]. Any unordered pair $u, v$ is called edge, and the vertices of this pair are adjacent from both $u$ and $v$.

Verbally, the projection $P(u)$ can be defined as a description of a subgraph of graph $G(V, E)$ containing sufficient information about all simple paths from the angle vertex $u \in V$ (projection base) to its other vertices which distances from $u$ do not exceed a given value. It is a sequence of levels, where subsets of adjacent vertices are generated at each level from which all vertices preceding the current one on the path from $u$ are excluded.

We will give a formal definition of the projection of an unweighted mixed graph in the form of a sequence of steps for its construction and demonstrate this by the example of the graph shown in Fig. 1 and the corresponding adjacency matrix (see Table 1).

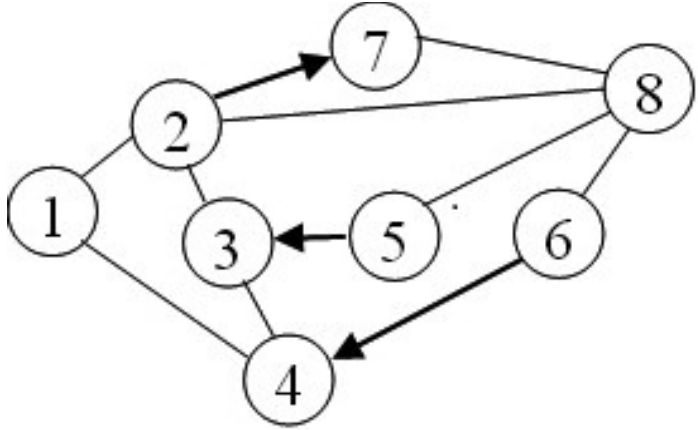

**Fig. 1** A simple mixed unloaded graph.

For reasons of better clarity, the adjacency matrix is used here as an initial one. Replacing it, e. g., with an adjacency list, which is more compact in the case of sparse graphs, will not require additional efforts due to the bijectivity of these descriptions.

**Table 1** Adjacency matrix $A$ of graph $G(V, E)$ shown in Fig. 1

| $v_i \backslash v_j$ | 1 | 2 | 3 | 4 | 5 | 6 | 7 | 8 |
|---|---|---|---|---|---|---|---|---|
| 1 | # | 1 | 0 | 1 | 0 | 0 | 0 | 0 |
| 2 | 1 | # | 1 | 0 | 0 | 0 | 1 | 1 |
| 3 | 0 | 1 | # | 1 | 0 | 0 | 0 | 0 |
| 4 | 1 | 0 | 1 | # | 0 | 0 | 0 | 0 |
| 5 | 0 | 0 | 1 | 0 | # | 0 | 0 | 1 |
| 6 | 0 | 0 | 0 | 1 | 0 | # | 0 | 1 |
| 7 | 0 | 0 | 0 | 0 | 0 | 0 | # | 1 |
| 8 | 0 | 1 | 0 | 0 | 1 | 1 | 1 | # |

Symbol "0" in the element (cell) $a_{ij}$ of matrix $A$ means absence of an arc or a vertex from vertex $v_i$ of the $i^{\text{th}}$ row to vertex $v_j$ of the $j^{\text{th}}$ column, where $a_{ij} = 1$, $a_{ji} = 0$ is indicative of the presence of arc $(v_i, v_j)$ from $v_i$ to $v_j$, or of edge $(v_i, v_j)$ if $a_{ij} = a_{ji} = 1$.

The projection $P(u)$ of graph $G(V, E)$ is a $k$-level structure, at the zero level of which the angle vertex $u \in V$ is located. The subset generated by this vertex $V_1 \subset V$ of vertices adjacent from $u$ is placed on the first level and contains all the vertices of its neighborhood $V_1 = N(u) = \{(v_{1,1})_w, (v_{1,2})_w, \cdots, (v_{1,|N(u)|})_w\}$.

Each vertex of the first level $v_{1,x} \in V_1$ generates, at the next level, a subset of $\{N(v_{1,1}) \setminus \{u\}\}$ adjacent vertices, excluding the vertex $u$ preceding vertices $v_{1,x} \in V_1$). The totality of these subsets determines all vertices of the $2^{\text{nd}}$ level, i.e., $V_2 = \{N(v_{1,1}) \setminus \{u\}, N(v_{1,2}) \setminus \{u\}, \cdots, N(v_{1,|N(u)|}) \setminus \{u\}\}$.

Each $i^{\text{th}}$ ($i > 1$) projection level is a set of neighborhoods of vertices of the $(i-1)^{\text{th}}$ level, excluding all the vertices preceding them, up to $u$. For example, for vertex $v_{(i-1),x}$, a subset of adjacent vertices $V_i(v_{(i-1),x})$ is created at the $i^{\text{th}}$ level without vertices preceding it on the path from $u$ to $v_{(i-1),x}$ of vertices $V_i(v_{(i-1),x}) = N(v_{(i-1),x}) \setminus (u - v_{(i-1),x})$. For vertex 1 of the graph shown in Fig. 1, the projection is as follows:

$$P(1) = 1^{(2^{(3^{(4)}, 7^{(8)}, 8^{(5,6,7)}, 4^{(3^{(2)})})})}.$$

The number of subsets generated at the $i^{\text{th}}$ level is equal to the number of instances of $(i-1)^{\text{th}}$ level vertices. Some of these subsets may be empty if $N(v_{(i-1),x}) \setminus (u - v_{(i-1),x}) = \emptyset$. [13, 16]

The relations of the immediate precedence and generation of vertices in the projection $P(u)$ are the relations of their adjacency from the preceding vertex[1]. Ordering the vertices by their immediate antecedents and generations from the angle vertex $u$ to the $j^{\text{th}}$ vertex of the $k^{\text{th}}$ level, we obtain a path $(u - v_{kj}) = (u, v_{1x}, \ldots, v_{kj})$ from vertex $u$ to vertex $v_{kj}$, while, in an unweighted graph, the number of level $k$ is equal to the length of a simple path from the angle vertex $u$ to any vertex of subset $V_{kj}$. If the path $(u - v_{kj})$ does not contain directed edges (arcs), then the reverse sequence of vertices of this path corresponds to the reverse path from $v_{kj}$ to the angle vertex $u$: $(v_{kj} - u) = \overline{(u - v_{kj})}$.

If the same vertex occurs $m$ times in the graph projection, then its instance closest to the base and located first on the left is considered original, and the remaining $(m-1)$ copies of it are called replicated, or replicas [17]. The appearance of replicas in the projection indicates the presence of cycles in the graph. The sum of the level numbers located in different branches of the original vertex and its replica, or any two replicas of the same vertex, determines the length of the corresponding cycle.

The projection $P(u)$ of a graph $G(V, E)$ is complete if it determines all vertices and all edges of this graph [18]. The vertex-complete projection of graph $G(V, E)$ contains all vertices of set $V$, but is not necessarily complete, whereas the edge-complete projection is always complete. The minimum number of levels of a vertex-complete projection is determined by the eccentricity $\varepsilon(u)$ of its base (angle vertex $u$), whereas, in a full projection, it can be one more if, at least, a pair of original vertices of level $k = \varepsilon(u)$ are adjacent.

Sufficiently detailed information about the properties of projections can be obtained also elsewhere (for example, [9, 10, 19]). However, in all these papers, we dwell upon projections of undirected unweighted graphs only. In [20] and here we complement these contributions and take into account the possible presence of not only undirected edges in the graph, but also oriented arcs. In Table 2 are the full-vertex projections of the mixed unweighted graph taken as an example (see Fig. 1).

The number of levels in the projections of graph $G(V, E)$ may vary depending on the problem being solved, but for the task of finding all shortest paths from one vertex, full-vertex projections of the graph are necessary; therefore, further, speaking of the projection $P(u)$, we endow it with the vertex completeness property by default.

---

[1] If there is an arc in a mixed graph directed from vertex $A$ to vertex $B$, then vertex $B$ is adjacent from vertex $A$, but from vertex $B$ vertex $A$ is not adjacent.

**Table 2** Projections of a mixed unweighted graph

$$P(1) = 1^{(2^{(3^{(4)},7^{(8)},8^{(5,6,7)},4^{(3^{(2)})})})}$$
$$P(2) = 2^{(1^{(4)},3^{(4)},7^{(8)},8^{(5,6,7)})}$$
$$P(3) = 3^{(2^{(1^{(4)},7^{(8)},8^{(5,6,7)})},4^{(1^{(2)})})}$$
$$P(4) = {(1^{(2^{(3^{()},7^{(8)},8^{(5,6,7)})})},3^{(2^{(1^{()},7^{(8)},8^{(5,6,7)})})})}$$

$$P(5) = 5^{(3^{(2^{(1,7,8)},4^{(1)})},8^{(2^{(1,3,7)},6^{(4)},7^{()})})}$$
$$P(6) = 6^{(4^{(1,3)},8^{(2,5,7)})}$$
$$P(7) = 7^{(8^{(2^{(1,3)},5^{(3)},6^{(4)})})}$$
$$P(8) = 8^{(2^{(1,3,7)},5^{(3)},6^{(4)},7^{()})}$$

Thus, the projection $P(u)$ of graph $G(V, E)$, constructed to a level $k$ equal to the eccentricity $\varepsilon(u)$, determines all simple paths from the angle vertex $u \in V$ to each vertex of the multiset $V'(k) = \bigcup_{i=1}^{k} V'_i, V'_i \subseteq V_i, |V'_i| \geq |V_i|$ of vertices of all levels, the distances to which, measured by the number of edges and arcs in the corresponding paths, do not exceed $k$.

To solve the SDSP problem of finding the shortest paths to one of the graph vertices, the concept of inverse projection $\overline{P}(u)$ may be needed. If, in the projection $P(u)$, the angle vertex $u$ is the initial one in all simple routes[2] with a length not exceeding its eccentricity $\varepsilon(u)$, then, in the inverse projection $\overline{P}(u)$, vertex $u$ is the final one in all simple routes limited in length by the inverse eccentricity[3] $\overline{\varepsilon}(u)$ from the remaining graph vertices $\overline{\varepsilon}(u) = \max_{v \in V} d(v, u)$. The inverse eccentricity $\overline{\varepsilon}(u)$ of vertex $u$ is the largest distance to this vertex out of the rest of its vertices. The number of levels in the direct $P(u)$ and reverse $\overline{P}(u)$ full-vertex projections is equal to the corresponding eccentricities $\varepsilon(u)$ and $\overline{\varepsilon}(u)$ of the angle vertex $u$. The inverse projections of the graph presented in Fig. 1 are shown in Table 3.

**Table 3** Inverse projections of a mixed graph

$$\overline{P}(1) = 1^{(2^{(3^{(4,5)},8^{(5,6,7)})},4^{(3^{(2,5)},6^{(8)})})},$$
$$\overline{P}(2) = 2^{(1^{(4)},3^{(4,5)},8^{(5,6,7)})},$$
$$\overline{P}(3) = 3^{(2^{(1^{(4)},8^{(5,6,7)})},4^{(1^{(2)},6^{(8)})},5^{(8^{(2,6,7)})})},$$
$$\overline{P}(4) = 4^{(1^{(2^{(3,8)})},3^{(2^{(1)},5^{(8)})},6^{(8^{(2,5,7)})})},$$

$$\overline{P}(5) = 5^{(8^{(2^{(1^{(4)},3^{(4)})},6^{()},7^{(2^{(1,3)})})})},$$
$$\overline{P}(6) = 6^{(8^{(2^{(1^{(4)},3^{(4,5)})},5^{()},7^{(2^{(1,3)})})})},$$
$$\overline{P}(7) = 7^{(2^{(1^{(4)},3^{(4,5)})},8^{(2^{(1,3)},5^{()},6^{()})})},$$
$$\overline{P}(8) = 8^{(2^{(1^{(4)},3^{(4,5)})},5^{()},6^{()},7^{(2^{(1,3)})})}$$

However, the above description is not always advantageous: specifying the goals set allows you to discard some details in the formal description of the object and simplify the achievement of these goals. In addition to the search problem discussed here, such a graph projection contains information about all paths, even those that obviously cannot be shortest. The solutions to SSSP and SDSP problems differ only in the use of direct and inverse projections, respectively. Therefore, the statements formulated and justified below, which make it possible to exclude from the projections a task that is redundant for solving the SSSP, will be correct for the SDSP task as well.

---

[2] We call a simple route in which, as in a simple chain, all vertices are pairwise distinct.
[3] The concepts of inverse eccentricity and inverse diameter make sense in oriented and mixed graphs and are first introduced here. In our opinion, these metrics can take their place among the metrics used in the analysis of such graphs, e. g., metrics of vertex centrality, etc.

# 3 Refining the projection and finding shortest paths

The lemma formulated below is based on the Bellman-Dijkstra optimality principle [2, 3], according to which any sections of the shortest path in positively weighted graphs are also shortest.

**Lemma 1.** *If the k-level projection $P(u)$ contains several instances of vertex $v$, then the shortest paths will be only those that pass through the instances of $v$ located at the lowest projection level.*

*Proof.* This lemma is a direct consequence of the Bellman-Dijkstra principle mentioned above. According to it, redundant information about all obviously non-shortest paths to vertex instances can be excluded from the initial projection $P(u)$, since the paths to such instances have a length greater than other paths determined by the projection being built, and, therefore, neither these paths nor their possible extensions are the shortest. In addition, the superimposing of the projection from such instances can also be stopped, since the paths to such instances have a length greater than other paths determined by the projection being built, and, therefore, neither these paths nor their possible extensions are the shortest. □

We show the application of Lemma 1 to the projection $P(4)$ of the demonstration graph. As a result, a refined projection $P'(4)$, from which vertex instances that do not satisfy the conditions of Lemma 1 were excluded, was obtained,. For a better visual comparison of the original $P(4)$ and the resulting $P'(4)$ projections, the excluded vertex instances in the latter are hidden (marked in gray):

$$P(4) = 4^{(1^{(2^{(3^{()},7^{(8)},8^{(5,6,7)})},3^{(2^{(1^{()},7^{(8)},8^{(5,6,7)})})})}} \to$$

$$P'(4) = 4^{1^{(2^{(3^{()},7^{(8)},8^{(5,6,7)})},3^{(2^{(1^{()},7^{(8)},8^{(5,6,7)})})}}} \to P'(4) = 4^{(1^{(2^{(7^{()},8^{(5,6)})},3^{(2^{(7^{()},8^{(5,6)})})})}}.$$

A refined $P'(u)$ projection can be obtained directly from the adjacency matrix, bypassing the stage of obtaining $P(u)$ if the procedure for its construction includes checking for the conditions defined by Lemma 1. Although the removal of vertices from $P(u)$ that are part of obviously non-shortest paths may lead to an increase in the number of levels in the projection $P'(u)$, in this case, the replication of vertices separated from $u$ at a distance greater than the shortest one will be excluded. Thus, the projection $P'(u)$ determines only the shortest paths from the angle vertex $u$ to all other graph vertices.

## 3.1 Building a refined $P'(u)$ projection

The ultimate goal of constructing the projection $P'(u)$ is to obtain a compact data structure from which all paths that are not shortest are excluded.

Formalize $P'(u)$ as a row and denote it $B(u)$; the cell number $j$ in it corresponds to the vertex $v_j$ of a simple path from $u$ to $v_j$, and the content $b_j$ is the number of the vertex preceding the vertex $v_j$ on the path from $u$ to $v_j$. Additionally, we use the following designation:

$A$ - graph adjacency matrix, $a_{i,j}$ - content of cell $(i, j)$ of the intersection of the $i^{\text{th}}$ row and the $j^{\text{th}}$ column of matrix $A$, $A(i)$ - set of vertices of the $i^{\text{th}}$ row of matrix $A$ adjacent from vertex $i$, $A(x) \equiv N(x)$ - neighborhood of vertex $x$;

$R_{u-v} = \{u, \ldots, v\}$ - set of vertices that make up a simple path from $u$ to $v$ and are forbidden to continue this path from vertex $v$;

$V_A$ - set of vertices, the shortest paths to which are known, the set of blocked for reading columns of the adjacency matrix $A$;

$V_k$ - multiset of the $k^{\text{th}}$ level vertices of projection $P'(u)$, $V_k(x)$ - set of vertices of the $k^{\text{th}}$ level $P'(u)$ generated by a $(k-1)^{\text{th}}$ level vertex $x$.

The process of obtaining a single-row refined projection $P'(u)$ is carried out step by step, where each step $k$ corresponds to the current level of its construction. And so:

**1) $k := 1$**. The set $V_1$ of vertices generated at the $1^{\text{st}}$ level is equal to the neighborhood $N(u)$: $V_1 := N(u) = A(u)$. Therefore, $b_{x \in V_1} := u$. For each vertex $x \in V_1$, we determine a set of $R_{u-x}$ vertices that are forbidden to continue through it: $R_{u-x} := u, x | x \in V_1$. We form a set of $V_A$ vertices, the shortest paths to which are found (including the angle vertex $u$ of the projection in it): $V_A := \{u\} \cup N(u)$. In accordance with Lemma 1, we block access to reading the columns of the adjacency matrix $A$ corresponding to the vertices from $V_A$. If the number of $|V_A|$ shortest paths found (blocked columns) coincides with the number of $|V|$ vertices in the graph, then the projection $P'(u)$ is made. Otherwise we proceed to point 2.

**2) $k := k + 1$** - increasing the number of the current projection level. For each vertex of the previous $(k-1)^{\text{th}}$ level, we determine a subset of vertices generated at the current $k^{\text{th}}$ projection level: $V_k(x \in V_{k-1}) := N(x) \setminus V_A \setminus R_{u-x}$. In this case, the neighborhoods $N(x \in V_{k-1}(x))$ are determined from the rows corresponding to the vertices $x \in V_{k-1}(x)$ of the rows of adjacency matrix $A$ modified in the previous steps (with columns blocked for reading corresponding to the shortest paths found earlier): $N(x \in V_{k-1}(x)) := N(x) \setminus V_A$. In $B$ cells, corresponding to the vertices $y \in V_k(x \in V_{k-1})$, we insert vertices $x$ preceding those of $y$: $b_y(y \in V_k(x \in V_{k-1})) := x$. A multiset of the vertices of the current level is obtained by combining these subsets: $V_k := \cup V_k(x | x \in V_{k-1}(x))$. For each vertex $x \in V_k$ of the current level $k$, we form a subset of $R_{u-x}$ vertices that are forbidden to continue the path from it: $R_{u-x} := R_{u-x} \cup \{x\} | x \in V_k$. If $|V_A| + |V_k| = |V|$, then we reset the column block ($V_A := \emptyset$) because the projection is completed. Otherwise we block the corresponding $V_k$ columns of the adjacency matrix $A$ for reading, modify $VA := V_A \cup Vk$ and repeat point 2.

Show this using the example of projection $P(4)$:

**1) $k := 1$**. $V(4) = V_1(4) := A(4) = \{1, 3\}$, $b_1 := 4$, $b_3 := 4$. $R_{4-1} := \{4, 1\}$, $R_{4-3} := \{4, 3\}$. $V_A := \{4, 1, 3\}$. As $|V_A| < |V|$ ($3 < 8$), the vertex completeness condition is not met, so we proceed to the fulfillment of point 2.

**2) $k := 2$**. Taking into account that $V_2 = \{V_2(x) | x \in V_1\} = \{V_2(x) | x \in \{1, 3\}\} = \{V_2(1), V_2(3)\}$, we get $V_2(1) := A(1) \setminus V_A = \{2, 4\} \setminus \{4, 1, 3\} = \{2\}$, $V_2(3) := A(3) \setminus V_A = \{2, 4\} \setminus \{4, 1, 3\} = \{2\}$. $V_2 := \{2\}$, $b_2 = \{1, 3\}$: this means that two paths of the same length lead to vertex 2 from vertex 4: $(4, 1, 2)$ and $(4, 3, 2)$. Designate them $(4-2)'$ and $(4-2)''$. Then $R'_{4-2} = \{4, 1, 2\}$, $R''_{4-2} = \{4, 3, 2\}$. $V_A := \{4, 1, 3\} \cup \{2\} = \{4, 1, 3, 2\}$. As $|V_A| < |V|$ ($4 < 8$), the vertex completeness condition is not met. We return to point 2.

**3)** $k := 3$. As two paths lead to vertex 2, then $V_3 = \{V_3(2_1), V_3(2_2)\} = \{A(2) \setminus V_A \setminus R'_{4-2}, A(2) \setminus V_A \setminus R''_{4-2}\}$. The content of the $2^{nd}$ row of matrix $A$, with blocked columns $V_A = \{4, 1, 3, 2\}$ taken into account is $A(2) \setminus V_A = \{1, 3, 7, 8\} \setminus \{4, 1, 3, 2\} = \{7, 8\}$. Therefore, multiset $V_3 := \{\{7, 8\} \setminus R'_{4-2}, \{7, 8\} \setminus R''_{4-2}\} = \{\{7, 8\}, \{7, 8\}\}$ and $V_A := V_A \cup \{7, 8\} = \{4, 1, 3, 2, 7, 8\}$. As $|V_A| < |V|$, then we return to point 2.

**4)** $k := 4$. $V_4 = \{V_4(7), V_4(8)\} = \{(A(7) \setminus V_A) \setminus R_{4-7}, (A(8) \setminus V_A) \setminus R4 - 8\} = \{\emptyset, \{5, 6\}\}$, $b_5 = b_6 := 8$. Since $V_A := V_A \cup \{5, 6\} = \{4, 1, 3, 2, 7, 8\} \cup \{5, 6\} = \{4, 1, 3, 2, 7, 8, 5, 6\}$, and $|V_A| = |V|$, the vertex completeness of projection $P'(4)$ has been reached, and its final form is shown in Table 4.

**Table 4** Refined $P'(4)$ projection of the graph under study

| $u_j$ | 1 | 2 | 3 | 4 | 5 | 6 | 7 | 8 |
|---|---|---|---|---|---|---|---|---|
| $u := 4$ | 4 | 1, 3 | 4 | # | 8 | 8 | 2 | 2 |

All refined projections obtained for the graph under study are summarized below in the shortest path matrix (see Table 5).

**Table 5** Shortest path matrix (SPM) of the graph under study

| $u_i \setminus u_j$ | 1 | 2 | 3 | 4 | 5 | 6 | 7 | 8 |
|---|---|---|---|---|---|---|---|---|
| 1 | # | 1 | 2, 4 | 1 | 8 | 8 | 2 | 2 |
| 1 | 2 | # | 2 | 1, 3 | 8 | 8 | 2 | 2 |
| 1 | 2, 4 | 3 | # | 3 | 8 | 8 | 2 | 2 |
| 1 | 4 | 1, 3 | 4 | # | 8 | 8 | 2 | 2 |
| 1 | 2,4 | 3, 8 | 5 | 3 | # | 8 | 8 | 5 |
| 1 | 4 | 8 | 4 | 6 | 8 | # | 8 | 6 |
| 1 | 2 | 8 | 2, 5 | 6 | 8 | 8 | # | 7 |
| 1 | 2 | 8 | 2, 5 | 6 | 8 | 8 | 8 | # |

## 3.2 Finding the shortest paths

The shortest paths from vertex $u$ to any of the graph vertices are determined by a sequential selection of vertices preceding each vertex of the desired path, starting from the final one. For example, for the shortest path $(4-5)_\delta$ we get: $b_{4,5} = 8$, $b_{4,8} = 2$, $b_{4,2} = \{1, 3\}$, $b_{4,1} = b_{4,3} = 4$, which corresponds to the two shortest paths: $(4-5)_\delta = \{(4, 1, 2, 8, 5), (4, 3, 2, 8, 5)\}$.

## 4 Complexity estimate

As can be seen from the previous section, the search for shortest paths consists of two stages. The stage of preliminary transformation of the adjacency matrix of a graph into its one-row refined projections. The space complexity $S_1$ of the first stage is determined by the space complexity of the graph adjacency matrix[4]: ($S_1 = O(n^2)$). At the second stage, the complexity will remain the same ($S_2 = O(n^2)$) for APSP or decrease to $S_2 = O(n)$ for SPSP, SSSP or SDSP problems, for which only one $B(u)$ row of $n$ rows of the SPM (shortest paths matrix) is sufficient.

Consider the asymptotic complexity $T_1$ of obtaining a refined projection $P'(u)$. Here, at each step, a set of vertices of the current level is determined. In this case, the projection is completed when it reaches the vertex completeness property, i.e., after filling all $n$ cells of the $P'(u)$-row; therefore $T_1 = O(n)$. Solving the search problem in the APSP-formulation (APSP - All-Pairs Shortest Path) will require constructing all $n$ projections of the graph. Therefore the asymptotic complexity of constructing the SPM will be quadratic: $T_1 = O(n^2)$.

Unlike the preliminary one, the asymptotic complexity of the shortest path sampling stage (SPSP problem) from the set of shortest paths already specified by the projection $P'(u)$ is determined by the distance $d(u,v)$ of vertex $v$ from $u$ and does not exceed the eccentricity of the source vertex $\varepsilon(u)$: $T_2(n) = O(\varepsilon(u))$. The eccentricity $\varepsilon(u)$ of any vertex $v \in V$ of graph $G(V,E)$ does not exceed its diameter $D(G)$, which depends not only on the size $n = |V|$, but on the number of $m$ and distribution of edges. So, for SPSP we can write: $T_2(n,m) = O(D(n,m))$.

Obviously, $D(n,m) = n$ is maximal in "linear graphs", in which $n \geq 3$ and the number of edges $m = n-1$, and this is the worst case for $T_2(n,m)$. In the best case, the diameter $D(n,m)$ is equal to one. When the number of edges $m = n(n-1)/2$, the graph becomes complete. It is clear that, for any intermediate values of $n \leq m < n(n-1)/2$, the diameter is $1 < D(G) < n$. Therefore, its asymptotics can be described as $D(G) = \Omega(1)$, $D(G) = O(n)$. Accordingly, the asymptotics of the $T_2$ stage of sampling from the projection $P'(u)$ of one shortest path is $T_2(n, m = n(n-1)/2) = \Omega(1)$, $T_2(n, m = n-1) = O(n)$. It follows from this that the complexity of stage $T_2$ for a graph of a given size $n$ has an inverse dependence on its (graph) density[5]. Even if the application software should include the $T_1$ stage, then the total time $T = T_1 + T_2 = \Omega(n)$ and $T = O(n)$, i.e, $T = \Theta(n)$ and then, also in this case, its dependence on the graph size is linear.

Compare the complexity of the proposed method with the Dijkstra algorithm traditionally used in SPSP and SSSP problems. It is known that the complexity of Dijkstra algorithm for these problems is the same when using the minimum heap to store and update the distances to the vertices: $T(n,m) = O(n+m)$. This is due to the fact that Dijkstra algorithm works for the $O(n+m)$ time. In each algorithm iteration, we add each vertex to the heap only once, and each edge is considered only once. Therefore, the overall algorithm complexity remains the same for both SPSP and SSSP[14]. Considering the above, we obtain: for a linear graph - $T(n, m = n-1) = \Omega(n)$ and for a $K_n$-graph — $T(n, m = n(n-1)/2) = O(n^2)$, which is $n$ times higher than the estimates of the asymptotic complexity of the method proposed in this paper. Note also that, in contrast, the complexity of Dijkstra algorithm directly depends on the density of the graph under study: finding the shortest path in a dense graph requires more time than in a sparse graph of the same size.

---

[4]When using the adjacency list as the initial data structure, the space complexity of the first stage of the search can be reduced to $S_1 = O(n+m)$.

[5]The density of a graph is usually determined by the ratio of the number of edges to the maximum possible number of edges in a graph with a given number of vertices.

The asymptotics of stage $T_2$ in the SSSP formulation, i.e., samples from the projection $P'(u)$ obtained at stage $T_1$ of all $n$ shortest paths from a given vertex of the graph $G(V, E)$, is determined by an $n$-multiple repetition of the sampling of a single path: $T_2(n, m = n(n-1)/2) = n\Omega(1) = \Omega(n)$ and $T_2(n, m = n-1) = nO(n) = O(n^2)$. It is easy to make sure that the introduction of the $T_1$ application software stage does not change the lower and upper bounds of the total $T = T_1 + T_2$ asymptotic complexity. These estimates coincide with the estimates of the complexity of Dijkstra algorithm. However, unlike it, as in the SPSP problem discussed above, they demonstrate a decrease in the complexity of finding shortest paths with an increase in the density of a graph of a given size.

Using the same approach as the one presented above for the SPSP and SSSP problems, it is easy to make sure that the solution of the APSP shortest path search problem based on the projective description of unweighted graphs is also more effective than the Floyd-Warshall or BFS algorithms traditionally used for this purpose, the complexity of which is $O(n^3)$ or $O(n(n + m))$, respectively[14].

# 5 Conclusion

The method for finding shortest paths, which is based on the use of projections to describe unweighted mixed graphs, was formulated in the present contribution. The asymptotic complexity analysis showed that this method will be quite effective in optimizing information exchanges, transportation, logistics and in other scientific and practical applications of the graph theory. One of the key features of this method, which distinguishes it from classical algorithms, is that its algorithmic complexity has an inverse dependence on the density of the analyzed graphs, which expands its application potentials. In addition, it should be noted that this method has a clear predisposition to parallelization, which provides additional advantages in scaling applications and the data they process.

Hopefully, the scientific research related to the projective description of graphs and their application in various theoretical and practical tasks will not only be useful for a wide range of experts, but their joint efforts in this field will undoubtedly lead to the further development of these results and to the expansion of their application scope.